# Directory Service Provided by DSCloud Platform

Lican Huang
Hangzhou Domain Zones Technology Co., Ltd
Hangzhou, China
Huang_lican@yahoo.co.uk , Licanhuang@zstu.edu.cn

*Abstract*—When there are huge volumes of information dispersing in the various machines, global directory services are required for the users. DSCloud Platform provides the global directory service , in which the directories are created and maintained by the users themselves. In this paper, we describe the DSCloud Platform directory service's functions, authorization, mounting users' local file systems , and usage scenery for education.

## I.  INTRODUCTION

There are huge amount information in Internet, some in Web Servers, cloud platforms, others in users' personal computers or mobiles. There are also many resources in Internet of Things. The information may be many types, including text, files, or even real-time sensing data.  On the other side, social activities of users also are classified into specific groups, such as Facebook , Linkedin,  Weichat, etc. Therefore, a directory service is beneficial for the users today.

   A directory service provides access to information in a directory. Unlike file systems in operating systems, global directory service for all users provides convenient access for global distributed information.

   Yahoo[1] and Amazon[2] provide directory service on the Internet.  However, these directory services are made by small number of specialists.  It is impossible to provide the satisfactory directory services for all the users.

    DSCloud platform[3] is a software whose final goal is to provide various services for global users based on semantic P2P networks[4,5,6,7,8,9,10], in which all information is classified into hierarchical domains. It consists of website and sP2P Server. The client user usually uses browser and sP2P Client to finish his tasks. For  DSCloud platform,  that all global users   share a global common directory is very important. Therefore, we implemented DSCloud platform to provides a global directory service by which the directories are created  and maintained by the users themselves. DSCloud platform organizes all directories using hierarchical tree from root directory.

   This paper presents a global directory service provided by DSCloud Platform, the authorization of the directory service , the functions for mounting users' local systems, and usage scenery for education.

## II.  GLOBAL DIRECTORY SERVICE PROVIDED BY DSCLOUD PLATFORM

   DSCloud Platform organizes all categories into hierarchical directory (domain) tree starting from  root "ALL" as Fig. 1 shows. In the paper, the directory and domain all most mean the same thing.  To create new  directory, the first step is to locate the parent  directory in which we create the new one. There are three ways to locate the parent directory.

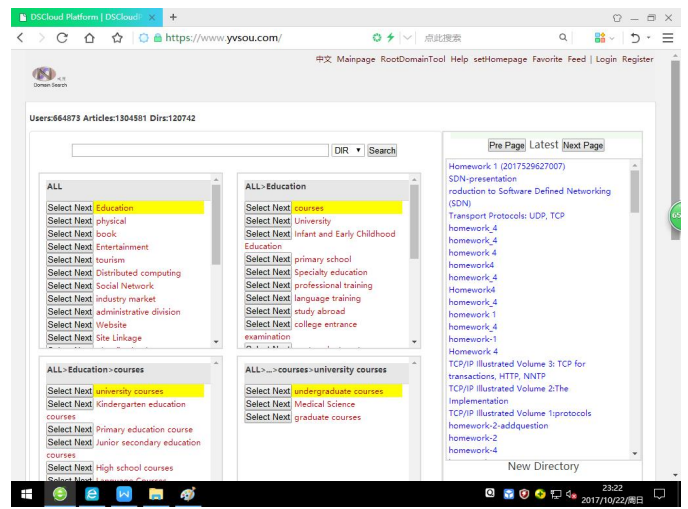

Fig.1. hierarchical directory (domain）tree

**DomainTool (Domain class )**

   At every  page of the platform , there is main menu containing "RootDomainTool menu".   RootDomainTool menu goes to DomainTool for root directory.  Every directory can go to its own DomainTool.   DomainTool has lots of functions including maintaining directory, managing user groups, authorization of the directory and other applications such as social communications and  online education, and so on .

   DomainTool shows current directory (current domain) and sub directories ( sub domains) as the  figure 2 shown.  We can step by step locate the desired directory  by  DomainTool from root via sub domain .

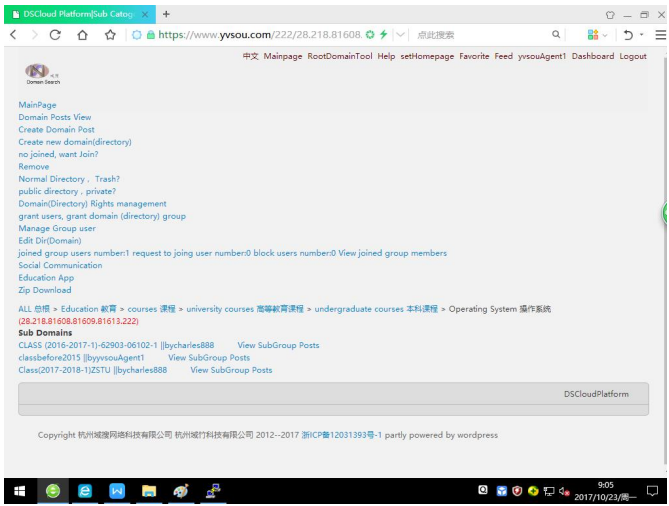

Fig.2. DomainTool

**Directory Navigator Bar (Domain Navigator Bar)**

Directory Navigator Bar is a directory bar, in which we can go to DomainTool and article list for any parent directory(as Fig. 3 shown).

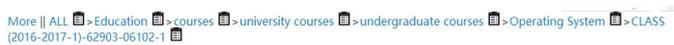

Fig.3. Directory Navigator Bar

By click the directory name (domain name) in Directory Navigator Bar, we can go to current DomainTool, by click the icon aside the directory name (domain name) in Directory Navigator Bar, we can go to the article list of the current domain. As the above figure, when click the blue words, it goes into the related domain tool. For example, when we click "Operating System", it will go to DomainTool for operating System directory.

**Search Box**

In main page of the platform, there is Search Box, by which we can quickly locate the desired directory( see Fig.4).

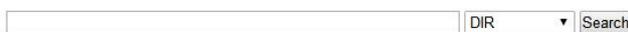

Fig.4. Search Box

The input of Search Box contains search words and option selection.

if search words contains " and " ("and" left side and right side have one space each), the target contents must satisfy these split words at the same time. For example, if using " protocol and course " to search directory, the target directory navigator bar must contains "protocol" and "course" at the same time.

If select option is "DIR", search the related directory, and display the contents as directory navigator bar, and help you quickly find the desired directory. If select option is "KEY", search the related words in article title and abstract, and display the article URL link with directory navigator bar. If select option is "My DIR", search the related directory which you created, and display the contents as directory navigator bar, and help you quickly find the desired directory. If select option is "MY KEY", search the related words in article title and abstract which you published, and display the article URL link with directory navigator bar. If select option is "MY ALL DIR", search all directories which you have created, and display the contents as directory navigator bar, and help you quickly find the desired directory.

## III. AUTHORIZATION OF DSCLOUD PLATFORM DIRECTORY SERVICE

Because DSCloud Platform aims to provide services to thousand millions of users, how to control the authorization of the directories is very important. In DSCloud Platform, the user who creates the directory is the creator and owner of the directory. The owner can delete the directory when there are no joined users and no published articles in this directory. The owner can also trash the directory. The owner can set the authorization of the directory created by himself(see Fig.5).

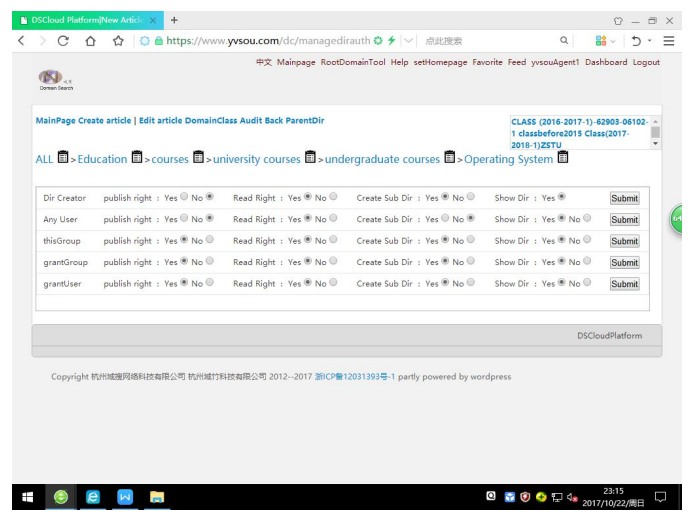

Fig.5. Setting Authorization of directory

The authorization schema is a matrix in which column is as user's role and row as the authorization. In Fig.5, we defined five roles as DirCreator, thisGroup, grantGroup, grantUser, and Any User, and four rights as publish, Read, Create Sub Dir and Show Dir.

Publish right means that the user who has the right can publish his articles( as well as attached files) in this directory. Read right means that the user who has the right can read the articles in this directory. Create Sub Dir right means that the user who has the right can create new sub directory in this directory, Show Dir right means that the user who has the right can see this directory name.

DirCreator is the role the user has who creates the directory. thisGroup is the role the user has who joined this directory(domain, sometimes we called as group) group. grantGroup is the role the user has who joined the group granted by the owner. grantUser is the role the user has who has been granted by the owner. Any User is the role for any login users.

The DSCloud Platform checks the user's right for these five roles. If any role among these fives with the right, and the user has the role, then the user has this right.

The owner of the directory grants the user of this directory role as Fig. 6 and Fig.7 shown.

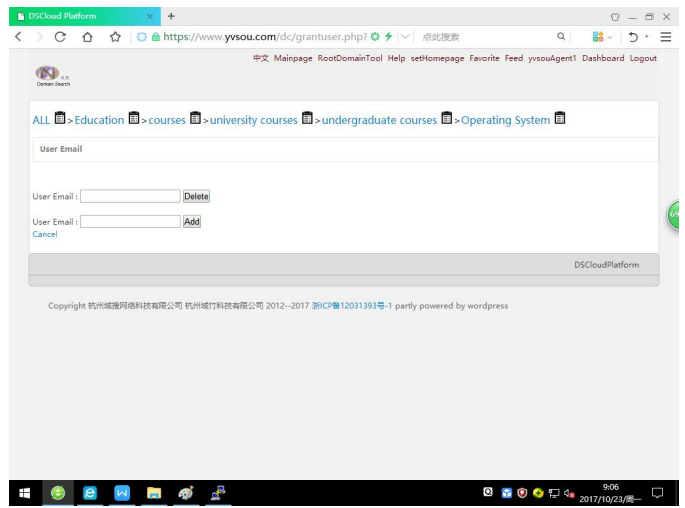

Fig.7. Grant user as the grantUser role for the directory.

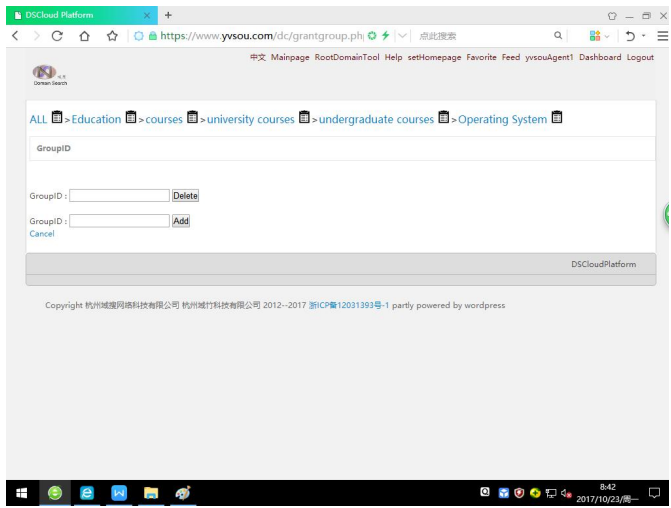

Fig.6. Grant group the grantGroup role for the directory.

## Group Management

The user who joined the group will has some rights. DSCloud Platform has the function for group management. The group has two types, public and private. Any user can join public group, but for private group, the owner is required to audit the user's application for joining the group.

DSCloud Platform can Set Directory Public or Private and manage group users by black, delete the group users.

The user can join Group by locating the related directory(domain), opening current domaintool (domain class), if "no joined, want join?", it means you have not joined the group. If you want to join, click the link. If the directory is public, then you join the group; however, if the directory is private, you just apply for joining the group, which will judge by the creator of the directory who can permit or refuse your application.

## IV. MOUNT USERS' LOCAL FILESYSTEMS

The user's local file systems can be mounted into the directory. We have developed the prototype for this function. The user's computer must download a client sP2P software, which communicates with sP2PServer and user's browser. The user machine create share local file directory, and binds its account into the directory, and using NAT technologies, when other machine browsers the directory, the use's local file directory will be shown in the browser at the same time. We hope this function will be published at this year.

## V. USAGE SCENERY

We here describe the usage scenery using the Platform to assist teaching university courses. The first step is register and login the web site. And then create related course directory. For example , we locate "Operating System " directory , and we create new directory "Class(2017-2018-1)ZSTU", and in this directory , we create help, homework, materials, notice , lecture materials, student homework directory, Exam and Test and lecture notes.

You can choose to set these groups public or private , set the authorization of these groups according to your requirements.

And ask students to create their own directory in the student homework directory , in which the student submits his homework.

We have use the Platform to assist teaching "Operating Systems", "Overview of computer sciences", "Modern Operating Systems", "Advanced Operating Systems", "Protocols and Implementations", "English Essentials", "Databases"; "Algebra"; "probability theory", " Chinese Poem", etc. We get all good evaluations from all teachers and students.

## VI. CONCLUSIONS

This paper presents a global directory service provided by DSCloud Platform. The directory service provide all users to create new directories and maintains their own directories. It also can set authorization policies for other users. The service can mount user's local file systems by using NAT technologies. We got good evaluations for usage scenery in education.

## ACKNOWLEDGMENT

The paper is supported by the project "Hangzhou Qinglan Plan--Scientific and technological creation and development (No.20131831K99" of Hangzhou scientific and technological committee. We thanks for Dr. Junhong Zheng, Dr. Junfu Ma, Dr. Min Zhao, Dr. Daolei Liang from Zhejiang Sci-Tech University, Prof. Shaozhong Zhang from Zhejiang Wanli University and other teachers and students to use DSCloud Platform to assist teaching their courses. The software copyrights is owned by Hangzhou Domain Zones Technology Co., Ltd. By agreements, Chinese patent applied is owned by Hangzhou Domain Zones Technology Company.